\documentclass[12pt]{article}
\usepackage{amssymb}
\usepackage{amsfonts}
\usepackage{amsmath}
\usepackage[mathscr]{eucal}
\usepackage{amssymb}
\usepackage{amsthm}
\usepackage{mathrsfs}
\usepackage{bbold}
\usepackage{bm}
\usepackage{graphicx}
\usepackage{caption}
\usepackage[english]{babel}
\usepackage[T2A]{fontenc}
\usepackage[utf8]{inputenc}
\usepackage{cite}
\usepackage{xcolor}

\newcommand{\be}{\begin{equation}}
\newcommand{\ee}{\end{equation}}
\newcommand{\bea}{\begin{eqnarray}}
\newcommand{\eea}{\end{eqnarray}}

\newcommand{\lb}{\label}
\newcommand{\p}[1]{(\ref{#1})}
\def\theequation{\arabic{section}.\arabic{equation}}

\usepackage{MnSymbol}

\newcounter{rown}

\begin{document}

\begin{titlepage}
\vspace*{0.3cm}

\begin{center}
{\LARGE\bf On BRST Lagrangian formulation for massive}\\

\vspace{0.5cm}

{\LARGE\bf higher-spin fields in $4D$ Minkowski space}

%\vspace{0.5cm}

%{\LARGE\bf improvements and simplifications}

%
\vspace{1.5cm}

{\large\bf I.L.\,Buchbinder$^{1,2,3}$,\,\, S.A.\,Fedoruk$^1$,\,\,  V.A.\,Krykhtin$^{2,4}$}

\vspace{1.5cm}

\ $^1${\it Bogoliubov Laboratory of Theoretical Physics,
Joint Institute for Nuclear Research, \\
141980 Dubna, Moscow Region, Russia}, \\
{\tt buchbinder@theor.jinr.ru, fedoruk@theor.jinr.ru}

\vskip 0.4cm

\ $^2${\it Department of Theoretical Physics,
Tomsk State Pedagogical University, \\
634041 Tomsk, Russia}, \\
{\tt joseph@tspu.ru, krykhtin@tspu.ru}

\vskip 0.4cm

\ $^3${\it National Research Tomsk State  University, 634050 Tomsk, Russia}

\vskip 0.4cm

\ $^4${\it Tomsk Polytechnic University, 634050 Tomsk, Russia}

\end{center}

\vspace{1.1cm}

\nopagebreak

\begin{abstract}
\noindent
We give a brief overview of the BRST approach to the gauge invariant Lagrangian formulation for free massive higher-spin bosonic fields focusing on two specific aspects. First, the theory is considered in four dimensional flat space in terms of spin-tensor fields with two component undotted and dotted indices. This leads to a significant simplification of the whole approach in comparison with the one where the fields with vector indices were used, since now there is no need to introduce a constraint responsible for the traces of the fields into the BRST charge. Second, we develop an extremely simple and clear procedure to eliminate all the auxiliary fields and prove that the BRST equations of motion identically reproduce the basic conditions for irreducible representations of the Poinc\'are group with a given mass and spin. Similar to the massless theory, the final Lagrangian for massive higher-spin fields is formulated in triplet form. The BRST formulation leads to a system of fields that are clearly subdivided into the basic spin $s$ field, Zinoviev-like auxiliary fields, Singh-Hagen-like auxiliary fields, and special BRST auxiliary fields. The auxiliary fields can be partially eliminated by gauge fixing and/or using the equations of motion. This allows one to obtain formally different (with different numbers of auxiliary fields) but equivalent Lagrangian formulations.

\end{abstract}

\vspace{1cm}


\phantom{Keywords: }

\newpage

\end{titlepage}

\setcounter{footnote}{0}
\setcounter{equation}{0}

\section{Introduction}
Higher spin field theory is one of the actively developing trends in modern theoretical physics (see e.g, the reviews \cite{revVas,revBCIV,revFT,revBek,DidSk,reviewsV,revBekSk,Snowmass,Ponomarev}).
Recently, there has been renewed interest in studing various aspects of the Lagrangian formulation for massive higher-spin fields (see, e.g., the papers \cite{OCHI,Lin,KUZ1,SkTs,KUZ2,DELP1,DELP2,DELP3} and references therein).

Lagrangian formulation of free massive arbitrary spin field theory in Minkowski space has been developed in the pioneer works by Singh and Hagen
\cite{Sing1,Sing2}.\footnote{Lagrangian formulation of a free massive field with an arbitrary integer or half-integer spin has also been considered by Schwinger in the monograph \cite{Schwinger} within the framework of the multispinor formalism; however, such an approach has not been developed for a long time.}
As was first pointed out in the work by Fierz and Pauli \cite{FP}, a local Lagrangian formulation of the free massive bosonic field with spin $s>1$ cannot be constructed in terms of this field alone, but requires a certain set of auxiliary fields of lower spins.
This means that the Lagrangian must have a very special structure. The corresponding equations of motion must lead to vanishing auxiliary fields together with the conditions determining the irreducible representation of the Poinc\'{a}re group with a given mass and spin. It is precisely this Lagrangian formulation that was constructed in \cite{Sing1}, although for some specific spins this had been done earlier. In general, it was established that the Lagrangian describing the dynamics of a free field with spin $s$ must include at least auxiliary fields with spins $s-2,\,s-3,\,\ldots,\,0.$ It should be noted that the procedure for eliminating auxiliary fields from the equations of motion, presented in \cite{Sing1}, is in general extremely cumbersome and is associated with rather nontrivial combinatorics.

The next important step in studying the structure of the Lagrangian formulation of a free massive field of arbitrary spin was made by Zinoviev \cite{Zin1,Zin2,Zin3}, where additional auxiliary St\"{u}ckelberg fields were introduced, which provided a gauge-invariant Lagrangian formulation and led to certain simplifications compared to the original formulation in \cite{Sing1}. The gauge invariant formulation was further developed by different authors (see e.g. \cite{Metsaev1,Metsaev2,Buch1,Buch2,Buch3,Buch4} and most detailed paper \cite{Zin4} and references therein).\footnote{This paper is a brief review of the BRST approach to the Lagrangian formulation of free massive fields of arbitrary integer spin. The issues related to massless fields, to fermion fields, to supersymmetric higher-spins and to the description of the interaction of higher-spin fields are undoubtedly also interesting but are beyond the scope of this review.}

The problem of constructing a gauge-invariant Lagrangian formulation of massive higher-spin fields with the required number of auxiliary fields, including the necessary St\"{u}ckelberg fields, is automatically solved within the BRST approach to higher-spin field theory \cite{Buch1,Buch2,Buch3,Buch4}.\footnote{Note also the papers \cite{Pash,T1,T2} where the Lagrangian description for the massive higher-spin fields was obtained by dimensional reduction of the massless theory in the BRST formalism. A similar reduction was considered in the later more detailed paper \cite{KUZ2}.} The central object in this approach is the BFV (Batalin, Fradkin, Vilkovisky) charge $Q$, also called the canonical BRST charge. By construction, the charge $Q$ is a Hermitian nilpotent operator acting in the auxiliary Fock space of vectors $\vert \Phi \rangle$ containing higher-spin fields and ghost fields. The free Lagrangian is constructed as ${\cal L} \sim \langle \bar\Phi \vert Q \vert \Phi \rangle$. Since the charge $Q$ is Hermitian, such a Lagrangian is automatically real. The corresponding equations of motion are written in the form $Q\vert \Phi \rangle = 0$. Due to the nilpotency of the BRST charge, both the Lagrangian and the equations of motion are invariant under the gauge transformations $\vert \Phi' \rangle = \vert \Phi \rangle + Q\vert \Lambda \rangle$ with the parameters $\vert \Lambda \rangle$ belonging to the same Fock space as the vectors $\vert \Phi \rangle$. The most important final step of the approach under discussion consists in proving that the equations of motion $Q\vert \Phi \rangle = 0$ automatically yield, as direct consequences, the conditions defining an irreducible representation with a given mass and spin. The latter circumstance ensures the correctness of the entire approach.\footnote{It is worth pointing out that this circumstance must be proved in any approach to higher-spin Lagrange formulation. In particular, it was one of the main results in the paper \cite{Sing1}.}

The BRST approach has two obvious advantages over the conventional approach, which directly uses higher-spin fields.\footnote{Power of the BRST approach in formulating nonlinear equations for supersymmetric higher-spin theory was emphasized in a recent paper \cite{Vas25}.} The Lagrangian obtained in this way contains all the necessary auxiliary fields by construction. Indeed, as is well known, beginning with \cite{FP}, the Lagrangian formulation of a higher-spin field cannot be constructed without using auxiliary fields. Since in the BRST approach the Lagrangian consistent with the conditions of the irreducible representation is presented explicitly, this automatically means the presence of the necessary auxiliary fields. In addition, in the BRST approach, the Lagrangian is gauge invariant. However, gauge invariance in massive theories is achieved by using suitable St\"{u}ckelberg fields. It follows that the Lagrangian constructed in the BRST approach automatically contains the necessary St\"{u}ckelberg fields. Therefore, the BRST approach to the theory of massive higher-spin fields is closely related to Zinoviev's gauge invariant Lagrangian formulation for massive higher-spin fields \cite{Zin1,Zin2,Zin3,Zin4}.

In this paper, we revise and essentially modify some aspects of the BRST approach to free massive integer higher-spin field theory in four-dimensional Minkowski space. Unlike the previous papers \cite{Buch1,Buch2,Buch3}, where fields with vector indices were considered, we propose here, similar to the papers \cite{BKout,BFIK-24}, that it more convenient to use, as fields, the irreducible representations of the Lorentz group with dotted and undotted two component indices. Such fields $\varphi_{\alpha(s)}^{\dot{\beta}(s)}=\varphi_{(\alpha_1\alpha_2 \ldots \alpha_s)}^{(\dot\beta_{1}\dot\beta_{2} \ldots \dot\beta_{s})}$ automatically satisfy the tracelessness condition that leads to a significant simplification of the whole construction. There is no need to introduce a constraint responsible for the field trace into the BRST charge. The second aspect concerns the proof that the BRST equation of motion identically reproduces the conditions defining the irreducible representation with a given mass and spin. The previous proofs \cite{Buch1,Buch2} were a bit long and very technical. Here we present an amazingly simple and clear derivation of the above conditions from the BRST equations of motion. Unlike the proof of these conditions in \cite{Sing1}, our proof uses only simple relations for annihilation and creation operators $[b,b^{+}]=1$, $b \,|0\rangle=0$ and avoids tedious combinatorics. The final gauge invariant Lagrangian is completely given in terms of the Fock space vector triplet including the basic higher-spin field, Singh-Hagen-like auxiliary fields, Zinoviev-like gauge auxiliary fields, and additional specific BRST auxiliary fields. Partial elimination of auxiliary fields due to their equations of motion and/or gauge fixing leads to formally different (e.g., involving different numbers of auxiliary fields) but certainly equivalent Lagrangian formulations.

The paper is organized as follows. Section\,2 is devoted to the formulation of free massive integer higher-spin fields in terms of the Fock space, the derivation of second class constraints, the  definition of a field of a given mass and spin, and the conversion of second class constraints into first class ones. In Section\,3 we develop the BRST construction, find the equations of motion and the gauge invariant Lagrangian in the extended Fock space, prove that these equations of motion yield correct conditions on the fields with a given mass and spin and describe the field contents of the theory under consideration. Section\,4 is a summary of the results. In the Appendix we present the resulting Lagrangian in terms of component fields including a full set of auxiliary fields following from the BRST formulation.

\setcounter{equation}0
\section{Description of free massive higher-spin fields in the Fock space}

\subsection{Fock space associated with massive higher-spin fields}
The conventional description of the Poinc\'are group irreducible representations with a given mass $m$  and a given spin $s$ is realized on totally symmetric tensor fields $\varphi_{m_1 \ldots m_s}=\varphi_{(m_1 \ldots m_s)}$ with $s$ Lorentz indices,   subject to the conditions
\footnote{The space-time metric is $\eta_{mn} = \mathrm{diag}(-1, +1, +1, +1)$.}
\be
\label{condition}
(\partial^m\partial_m - m^2)\varphi_{m_1 \ldots m_s}=0\,, \qquad \partial^{m_1}\varphi_{m_1 \ldots m_s}=0\,,
\qquad \eta^{m_1m_2}\varphi_{m_1m_2 \ldots m_s}=0.
\ee
However, in four dimensions it is much more convenient to convert each vector index into a pair of two-component indices $\alpha\,\, (\alpha=1,2)$ and $\dot\alpha\,\,(\dot\alpha = \dot{1},\dot{2})$\footnote{See the details in \cite{BK}.} and work with fields $\varphi^{\dot\beta(s)}_{\alpha(s)}(x)$ of the form
\be
\label{twocomp}
\varphi^{\dot\beta(s)}_{\alpha(s)}= (\sigma^{m_1})_{\alpha_1}{}^{\dot{\beta}_1} \ldots (\sigma^{m_s})_{\alpha_s}{}^{\dot{\beta}_s}\varphi_{(m_1 \ldots m_s)}.
\ee
In terms of the fields $\varphi^{\dot\beta(s)}_{\alpha(s)}(x)$ the last of the conditions (\ref{condition}) is fulfilled automatically and the rest conditions are rewritten as follows:
\be
\label{condition1}
(\Box - m^2)\varphi^{\dot\beta(s)}_{\alpha(s)}(x)=0\,, \qquad
\partial^{\alpha}_{\dot{\beta}}\varphi^{\dot\beta(s)}_{\alpha(s)}(x) =0\,.
\ee
Here $\displaystyle \Box=\partial^2=\partial^m\partial_m=\frac{1}{2}\,\partial_{\alpha}^{\dot\beta}\partial^{\alpha}_{\dot\beta}$
and $\partial^{\alpha}_{\dot{\beta}}\varphi^{\dot\beta(s)}_{\alpha(s)} :=
\partial_{\dot\gamma}^{\gamma}\varphi^{\dot\gamma\dot\beta_1\ldots\dot\beta_{s-1}}_{\gamma\alpha_1\ldots\alpha_{s-1}}$.

Let us introduce the Fock space associated with the fields $\varphi^{\dot\beta(s)}_{\alpha(s)}(x)$. For this purpose, one defines the bosonic creation
$\bar{c}_{\dot{\alpha}},\,\,c^\alpha$ and annihilation $\bar{a}^{\dot{\alpha}},\,\,a_\alpha$ operators subject to the conditions
\be\label{ac-vac}
\langle0|\bar{c}_{\dot{\alpha}}=\langle0|c^\alpha=0\,,
\qquad
\bar{a}^{\dot{\alpha}}|0\rangle=a_\alpha|0\rangle=0\,,
\qquad
\langle0|0\rangle=1
\ee
and commutations relations
\be
[\bar{a}^{\dot{\alpha}},\bar{c}_{\dot{\beta}}]
=\delta^{\dot{\alpha}}_{\dot{\beta}}\,,
\qquad
[a_\alpha,c^\beta]=\delta_\alpha^\beta\,.
\ee
The rest commutation relations vanish. The Hermitian conjugation for these operators looks like:
 \be
(a_\alpha)^+=\bar{c}_{\dot{\alpha}}\,,\qquad
(\bar{a}^{\dot{\alpha}})^+=c^\alpha\,.
\ee

Massive higher-spin Fock space vectors are defined as follows:
\be \label{GFState}
|\varphi_{s}\rangle=\frac{1}{s!}\,\varphi^{\dot{\beta}(s)}_{\alpha(s)}(x)\,c^{\alpha(s)}\,\bar{c}_{\dot{\beta}(s)}|0\rangle\,.
\ee
The dual (conjugate) vectors are written in the form
\be \label{GFState-a}
\langle\bar{\varphi}_s|=\frac{1}{s!}\,\langle 0|\,\bar{a}^{\dot{\alpha}(s)}\,a_{\beta(s)} \bar{\varphi}^{\beta(s)}_{\dot{\alpha}(s)}\,.
\ee

\subsection{Constraints for massive higher-spin Fock space vectors}
In this subsection we reformulate the conditions of the irreducible representation (\ref{condition1}) as the conditions on the Fock space vectors
$\vert\varphi_{s}\rangle$. Let us introduce the operators
\be
\label{l_0}
{l}_0=\Box-m^2\,,\qquad
{l}=\partial_{\alpha\dot{\beta}}a^\alpha\bar{a}^{\dot{\beta}}\,,\qquad
{l}^+={}-c^\beta\partial_{\beta\dot{\alpha}}\bar{c}^{\dot{\alpha}}\,.
\ee
The only non-vanishing commutator  for these operators has the form
\be
\label{THEequations2}
[{l}^+,{l}]=K{l}_0+m^2K\,,
\ee
where
\be\lb{KNbN}
K=N+\bar{N}+2\,,\qquad N=c^\alpha a_\alpha\,,\qquad
\bar{N}=\bar{c}_{\dot{\alpha}}\bar{a}^{\dot{\alpha}}\,.
\ee
The commutation relation including (\ref{l_0}) and (\ref{KNbN}) is written as follows:
\be\label{N-l}
[{l},N]={}{l}\,,\qquad[{l},\bar{N}]={l}
\,,\qquad [{l}^+,N]={}-{l}^+\,,\qquad [{l}^+,\bar{N}]={}-{l}^+\,,
\ee
\be\label{K-l}
[{l},K]={}2{l}\,,\qquad
[{l}^+,K]={}-2{l}^+\,.
\ee

Consider the subclass of the Fock space vectors $\vert{\varphi}_{s}\rangle$ subject to the conditions
\be
{l}_0|\varphi_{s}\rangle=0\,,\qquad
{l}|\varphi_{s}\rangle=0
\label{THEequations}
\ee
and show that they describe the irreducible representations with a given mass $m$ and spin $s$, where the spin takes the values $s=0,1, \ldots $. The first relation
in (\ref{THEequations}) evidently corresponds to mass irreducibility. To check spin irreducibility it remains to calculate the action of the second Casimir operator of the Poinc\'are group on the vectors $\vert\varphi_{s}\rangle$.

It is easy to see that the vectors $|\varphi_{s}\rangle$ satisfy the relations
\be
N|\varphi_{s}\rangle=s|\varphi_{s}\rangle\,,\qquad \bar N|\varphi_{s}\rangle=s|\varphi_{s}\rangle
\label{THEequations1a}
\ee
and hence, the following constraint automatically takes place
\be\label{THEequations1}
K|\varphi_{s}\rangle=(2s+2)|\varphi_{s}\rangle\,.
\ee

The anti-Hermitian translation operator
is $P_m=\partial_m$
and the anti-Hermitian Lorenz rotation operator $\mathcal{M}_{m n}$ has the form
\be
\lb{sn-3}
\mathcal{M}_{m n} = M_{m n} + \bar{M}_{m n} \,, \qquad
M_{m n} := c^{\alpha}\, (\sigma_{m n})^{\;\; \beta}_{\alpha} \,
a_{\beta} \,, \quad \bar{M}_{m n} := \bar{c}_{\dot{\alpha}} \,
(\tilde{\sigma}_{m n})^{\dot{\alpha}}_{\;\;\dot{\beta}} \,
\bar{a}^{\dot{\beta}} \,,
\ee
where $(\mathcal{M}_{m n})^\dagger={}-\mathcal{M}_{m n}$.
Then, the square of the Pauli-Lubanski operator
$\displaystyle
W_m={}-\frac12\,\epsilon_{mnkl}P^n \mathcal{M}^{kl} $ looks like
\be\label{PL-val}
W^mW_m \ = {}-(c\sigma^m\bar c)(a\sigma^n\bar
a)\,\partial_m\partial_n \ +\ \left[\frac{N}{2}\left(\frac{N}{2}+1
\right)+ \frac{\bar N}{2}\left(\frac{\bar N}{2}+1 \right)
+\frac{N\bar N}{2}\right]\partial^m\partial_m\,.
\ee
It is evident that the action of this operator on the vectors $|\varphi_{s}\rangle$, subject to \eqref{THEequations}, has the form
$W^mW_m=m^2\, s(s+1)$ where $\displaystyle s=\frac{N+ \bar{N}}{2}$\,. As a result, the constraints (\ref{THEequations}) define the irreducible representation of the
Poinc\'are group with a given mass $m$ and given spin $s$ in the Fock space.

To construct real Lagrangian within the framework of the BRST approach, we should have the Hermitian BRST charge. However, the BRST charge constructed on the basis of constraints (\ref{THEequations}) will be non-Hermitian, since the system of constraints (\ref{THEequations}) is not closed under the Hermitian conjugation. In principle, we can extend the system of constraints (\ref{THEequations}) adding the operator ${l}^+$. However, since the operators $l$ and ${l}^+$ do not commute, the extended system of constraints will form a  system of second class constraints and the standard construction of the BRST charge is not applicable. In the next subsection we will show how to convert a system of second class constraints into a system of first class constraints.

\subsection{Conversion of second class constraints}

Since the BRST charge is constructed on the basis of the first class constraints, we introduce a new equivalent system of first class constraints closed under Hermitian conjugation. Such a procedure assumes involving additional variables and is called the conversion of second class constraints \cite{FSh}, \cite{BF}.

The second class constraints ${l}$ and ${l}^+$ are formulated in terms of oscillators $\bar{c}_{\dot{\alpha}},\,\,c^\alpha$ and  $\bar{a}^{\dot{\alpha}},\,\,a_\alpha$. For conversion we add to these oscillators new bosonic oscillators
$b$ and $b^+$ subject to the standard commutation relations:
\be
[b,b^+]=1\,.
\label{b-com}
\ee
The converted constraints are defined as follows:
\be
\label{conver-op}
\ell_0={l}_0+{l}_0'\,,\qquad  \ell={l}+{l}'\,,\qquad \ell^+={l}^++{l}^{+\prime}
\ee
Here ${l}_0$, ${l}$, ${l}^+$ are the initial constraints \p{l_0} and  ${l}_0'$, ${l}'$, ${l}^{+\prime}$ are the additional parts dependent on additional annihilation and creation operators $b$ и $b^+$. The main principle of the conversion procedure states that the converted constraints unlike initial ones should form the first class constraint algebra.

In the case under consideration, the converted constraints are constructed as follows:
\begin{itemize}
\item
The mass constraint ${l}_0$ is not changed under conversion
\be
\label{conver-0}
\ell_0={l}_0\,.
\ee
\item
Additional parts $l'$, $l^{+\prime}$ are taken in the form:
\be
\label{prime-op}
{l}'=A_s(\mathrm{n})b\,,\qquad {l}^{+\prime}=b^+ A_s(\mathrm{n})\,,
\ee
where $A_s(\mathrm{n})$ is some function of the operator
\be
\label{n-op}
\mathrm{n}:=b^+ b\,.
\ee
It is assumed that this function can be represented as a series in the operator $\mathrm{n}$ \p{n-op}.
Thus, the operators \p{prime-op} are  conjugate to one another and are linear in $b$ and $b^+$ up to $A_s(\mathrm{n})$.
\item
The function $A_s(\mathrm{n})$ is found from the condition that the converted constraint algebra is closed.

\end{itemize}

Let us begin constructing converted constraints based on the above conditions. Since the converted constraints should contain the new operators
$b$ and $b^+$, we have to generalize the Fock space using the operator
$b^+$ and introduce the vectors
\footnote{We use the mass parameter $m$ in this expansion to explicitly distinguish between the massive and massless cases and to be able to have the limit $m\to 0$ at the level of states/fields. }
\be
|\phi_{s}\rangle=
\sum_{k=0}^{s}\frac{m^k}{\sqrt{k!}}
\, (b^+)^k
|\varphi_{s-k}\rangle\,,
\label{GFState-g}
\ee
where the vectors $|\varphi_{r}\rangle$ ($k=0\div s$) are defined as before \p{GFState}:
\be \label{GFState-g1}
|\varphi_{r}\rangle=
\frac{1}{r!}\,\varphi^{\dot{\beta}(r)}_{\alpha(r)}(x)\,c^{\alpha(r)}\,\bar{c}_{\dot{\beta}(r)}|0\rangle.
\ee
The vacuum is also extended,
\be
b|0\rangle =0\,,\qquad \langle 0|\,b^+ =0\,.
\label{b-vac}
\ee
The conjugate extended vector is written as follows:
\be \label{GFState-g-a}
\langle\bar{\phi}_{s}|=
\sum_{k=0}^{s}\frac{m^k}{\sqrt{k!}}
\,  \langle\bar{\varphi}_{s-k}|
(b)^k\,,
\ee
where $\langle\bar{\varphi}_s|$ is define like \p{GFState-a}:
\be \label{GFState-g1-a}
\langle\bar{\varphi}_r|=\frac{1}{r!}\,\langle 0|\,\bar{a}^{\dot{\alpha}(r)}\,a_{\beta(r)} \bar{\varphi}^{\beta(r)}_{\dot{\alpha}(r)}\,.
\ee

As we noted, the converted constraints have the form
\be
\label{el}
\ell = l+A_s(\mathrm{n})b\,,\qquad \ell^{+}= l^{+}+b^{+} A_s(\mathrm{n})\,.
\ee
They contain the operators $b$ and $b^+$ and  act in the Fock space of the vectors $|\phi_{s}\rangle$ (\ref{GFState-g}).
The commutator of the initial constraints $l$ and $l^+$ is given by relation (\ref{THEequations2}) where the second term in the right-hand side violates the closure of the generators algebra.
We will require that the commutator of the constraints $\ell$ and $\ell^{+}$ on the vectors $|\phi_{s}\rangle$ (\ref{GFState-g}) be closed on $\ell_{0}$ and the system of constraints on the above vectors be a first class one. To provide this condition, we postulate the commutator  $\ell$ and $\ell^{+}$ in the form
 \be
\label{THEequations2a}
\big[\ell^+,\ell\big]=K \ell_0+2m^2\big(S-s\big)\,,
\ee
where the operator
$S$ should preserve the operators \p{el}:
\be\label{K-L}
\big[\ell,S\big]=\ell\,,\qquad
\big[\ell^+,S\big]={}-\ell^+\,
\ee
 and act on the vectors $|\phi_{s}\rangle$ (\ref{GFState-g}) as follows
\be\label{K-phi}
S|\phi_{s}\rangle=s|\phi_{s}\rangle.
\ee
That is, the condition $S-s=0$ on the vectors $|\phi_{s}\rangle$ is satisfied.
Using the form of the operators $\ell$, $\ell^+$ \p{el} and conditions
\p{K-L}, \p{K-phi} one can find that the operator $S$ has the form
\be
\label{K-form}
S=\frac{1}{2}\,\big(N+\bar N\big) +\mathrm{n} = \frac{1}{2}(K-2)+\mathrm{n} \,,
\ee
where $\mathrm{n}$ is defined in \p{n-op}.
As a result, on the vectors $|\phi_{s}\rangle$ (\ref{GFState-g}) the commutator (\ref{THEequations2a}) reduces to
\be
\big[\ell^+,\ell\big]=K \ell_0.
\label{com-el}
\ee
All that remains is to find the function $A_s(\mathrm{n})$.

The form of the function $A_s(\mathrm{n})$ is dictated by the requirement to fulfil relation \p{THEequations2a}.
Using the relations
\be
b A_s(\mathrm{n})=A_s(\mathrm{n+1})b\,,\qquad
A_s(\mathrm{n})b=bA_s(\mathrm{n-1})\,,
\label{A-b-com}
\ee
\be
b^+ A_s(\mathrm{n})=A_s(\mathrm{n-1})b^+\,,\qquad
A_s(\mathrm{n})b^+=b^+A_s(\mathrm{n+1})\,,
\label{A-bp-com}
\ee
based on \p{b-com}, one gets
\be
\label{LL-com-a}
\big[b^+ A_s(\mathrm{n}),A_s(\mathrm{n})b\big]=
{}\mathrm{n}A_s(\mathrm{n}-1)A_s(\mathrm{n}-1)-
(\mathrm{n}+1)A_s(\mathrm{n})A_s(\mathrm{n})\,.
\ee
To get the commutator  \p{THEequations2a}, one puts
\be
A_s(\mathrm{n})=m \big(2s+2- \mathrm{n}\big)^{1/2}\,.
\label{A-val}
\ee
As a result, the operators \p{el} have the final form
\be
\label{new-op-f}
\ell \ = \ {l} \ + \ m \big(2s+2- \mathrm{n}\big)^{1/2} b\,,\qquad \ell^+ \ = \ {l}^+ \  + \  m b^+  \big(2s+2- \mathrm{n}\big)^{1/2}\,.
\ee

In the Fock space of the vectors $|\phi_{s}\rangle$ satisfying equation (\ref{K-phi}) the operator $S-s$ vanishes and we arrive at the following first class constraints:
\be
\label{op-dyn}
\ell_0\,,\qquad \ell\,,\qquad \ell^+\,.
\ee
The only nonzero commutator among these constraints has the form
\be
\label{THEequations2b}
\big[\ell^{+},\ell\big]=K \ell_0\,.
\ee
Let us emphasize, the situation in the massive higher-spin theory is completely analogous to the massless one (see, e.g., \cite{BFIK-24}). The only difference is the presence in the Fock space of one more creation operator $b^+$ in comparison with the massless case. It necessitates, unlike the massless case, to use  a tower of Fock space vectors with all spins from 0 to $s$ in the corresponding BRST construction, which is the reason for a large number of auxiliary fields in the massive theory.

Pay attention to that the operator \p{A-val} is well defined as a convergent power series.

\setcounter{equation}0
\section{Lagrangian formulation}
In this section we will derive the Lagrangian formulation for the theory under consideration and discuss the fundamental principal issue related with obtaining the conditions (\ref{THEequations}) that define the irreducible representations of the Poinc\'are group with a given mass and spin. The derivation of these conditions as direct consequences of the equations of motion is the justification of the correctness of the Lagrangian formulation for free higher-spin field theory.

\subsection{Equations of motion and Lagrangian in terms of extended Fock space}
The canonical BRST construction is realized in the extended Fock space where the fermionic ghost coordinate operators  $\eta_{0}$, $\eta$, ${\eta}^{+}$ and the corresponding ghost momenta $\mathcal{P}_{0}$, $\mathcal{P}^{+}$, $\mathcal{P}$ are added to the bosonic oscillators $c$, $\bar{c}$, $a$, $\bar{a}$, $b$, $b^+$. The ghosts $\eta_{0}$ and $\mathcal{P}_{0}$ are Hermitian, $\eta^ {+}_{0}=\eta_{0}$, $\mathcal{P}^{+}_{0}=\mathcal{P}_{0}$, whereas $(\eta)^{+}={\eta}^{+}$, $(\mathcal{P})^{+}=\mathcal{P}^{+}$. The only nonzero anticommutators for the ghost variables have the form
\be
\{\eta,\mathcal{P}^+\}
 \ = \ \{\mathcal{P}, \eta^+\}
 \ = \ \{\eta_0,\mathcal{P}_0\}
 \ = \ 1.
\label{ghosts}
\ee
The extended vacuum $|0\rangle$ is defined, besides \p{ac-vac} and \p{b-vac}, by the conditions
\be \lb{ghost-vac}
\eta|0\rangle \ = \ \mathcal{P}|0\rangle \ = \ \mathcal{P}_0|0\rangle \ = \ 0\,.
\ee
The ghost operators are characterized by the following ghost numbers:
$gh(\eta_0)= gh(\eta)=gh(\eta^+)=1$, $gh(\mathcal{P}_0)=gh(\mathcal{P})=gh(\mathcal{P}^+)=-1$.

The vectors of the extended Fock space are written in the form
\be
|\Phi\rangle \ = \
|\phi\rangle \ + \ \eta_0\mathcal{P}^+|\phi_1\rangle \ + \ \eta^+\mathcal{P}^+|\phi_2\rangle\,,
\label{extened vector}
\ee
and possess zero ghost number.

The extended Fock space for higher-spin field theory under consideration is obtained by extension of the vectors
\p{GFState-g} by the ghost operators. Therefore, the corresponding vectors in extended space look like
\be
\label{Phi-s}
|\Phi_s\rangle=
|\phi_s\rangle \ + \ \eta_0\mathcal{P}^+|\phi_{1(s-1)}\rangle \ + \ \eta^+\mathcal{P}^+|\phi_{2(s-2)}\rangle
\,,
\ee
where $|\phi_s\rangle$ are defined in \p{GFState-g}, \p{GFState-g1} and
\be \label{GFState-g-1}
|\phi_{1(s-1)}\rangle=
\sum_{k=0}^{s-1}\frac{m^k}{\sqrt{k!}}
\, (b^+)^k
|\varphi_{1(s-1-k)}\rangle\,,
\qquad
|\varphi_{1(r)}\rangle=
\frac{1}{r!}\,\varphi_{1,\,}{}^{\dot{\beta}(r)}_{\alpha(r)}(x)\,c^{\alpha(r)}\,\bar{c}_{\dot{\beta}(r)}|0\rangle\,,
\ee
\be \label{GFState-g-2}
|\phi_{2(s-2)}\rangle=
\sum_{k=0}^{s-2}\frac{m^k}{\sqrt{k!}}
\, (b^+)^k
|\varphi_{2(s-2-k)}\rangle\,,
\qquad
|\varphi_{2(r)}\rangle=
\frac{1}{r!}\,\varphi_{2,\,}{}^{\dot{\beta}(r)}_{\alpha(r)}(x)\,c^{\alpha(r)}\,\bar{c}_{\dot{\beta}(r)}|0\rangle\,.
\ee
As a result we arrive at a triplet of vectors $|\phi_{s}\rangle$, $|\phi_{1(s-1)}\rangle$, $|\phi_{2(s-2)}\rangle$ in the initial Fock space.

Let us turn to constructing the BRST charge. According to the general scheme, the BRST charge is derived with the help of first class constraints. Therefore, using the constraints $\ell_{0},\,\ell,\,\ell^{+}$, one obtains the following Hermitian operator:
\be\label{Q(0)}
\tilde Q \ = \ \eta_0 \ell_0 \ + \ \eta^+ \ell \ + \ \eta\,\ell^{+} \ + \
K\eta^+\eta\, \mathcal{P}_0 \,.
\ee
However,
\be
\label{Q-sq}
\tilde Q^2=
{}-2m^2\eta^+\eta\left(S-s\right)=
{}-2m^2\eta^+\eta\left(\hat S-s-1\right),
\ee
where the Hermitian operator $\hat{S}$ has the form
\be
\hat{S} \ = \ S+\eta^+\mathcal{P}+\mathcal{P}^+\eta  \ = \ \frac{1}{2}\left( N+\bar{N}\right) +\mathrm{n}+\eta^+\mathcal{P}+\mathcal{P}^+\eta\,
\label{op-NS}
\ee
and the operator $S$ is defined in (\ref{K-form}). The operator $\hat{S}$ acts in the extended Fock space and generalizes the operator $S$ that defines the true massive higher-spin field due to relation
(\ref{K-phi}). To define subspace of true massive higher-spin fields in the extended Fock space, it is natural to impose the condition
\be\lb{S-eq}
\hat S\,|\Phi_s\rangle=s\,|\Phi_s\rangle \,.
\ee
Vectors \p{Phi-s} of the extended Fock space obey precisely this condition.
However, according to (\ref{Q-sq}), the operator (\ref{Q(0)}) will not be nilpotent in subspace of the vectors $|\Phi_s\rangle$ subject to relation (\ref{S-eq}).
The reason for the loss of nilpotency may be due to the use of not quite suitable constraints. Instead of constraints $\ell$ and $\ell^{+}$, one introduces new constraints according to the rule
\be
\label{h-new-op}
\begin{array}{rcl}
\hat\ell &=&{l} \ + \ A_{s-1}(\mathrm{n})b \ = \ {l} \ + \ m \big(2s- \mathrm{n}\big)^{1/2} b\,,\\ [6pt]
\hat\ell^+&=&{l}^+ \ + \ b^+ A_{s-1}(\mathrm{n}) \ = \ {l}^+ \  + \  m b^+  \big(2s- \mathrm{n}\big)^{1/2}\,.
\end{array}
\ee
The constraint $\ell_{0}$ does not change.
The commutator of the constraints (\ref{h-new-op}) has the form
\be
\label{THEequations2ab}
\big[\hat\ell^+,\hat\ell\big] \ = \ K \ell_0+2m^2\big(S-s+1\big)\,.
\ee

Let us construct the BRST charge on the basis of the constraints $\ell_{0},\, \hat\ell,\, \hat\ell^+$ and show that it will be Hermitian and nilpotent in subspace of vectors subject to (\ref{S-eq}). The general definition of the BRST charge allows one to introduce the operator
\be\label{Q(r)}
Q \ = \ \eta_0 \ell_0 \ + \ \eta^+ \hat\ell \ + \ \eta\,\hat\ell^{+} \ + \
K\eta^+\eta\, \mathcal{P}_0 \,.
\ee
Expression (\ref{Q(r)}) leads to
\be\label{Q-sq-r}
Q^2={}-2m^2\eta^+\eta\left(S-s+1\right)={}-2m^2\eta^+\eta\left(\hat S-s\right)
\ee
Thus, the operator (\ref{Q(r)}) is nilpotent on the vectors \p{Phi-s} under the condition \p{S-eq}: $Q^2|\Phi_s\rangle=0$.

The equation of motion for a free bosonic massive higher-spin field is postulated as follows:
\begin{eqnarray}\label{eqQ}
Q\,|\Phi_s\rangle \ = \ 0.
\end{eqnarray}
It is worth emphasizing that this postulate is not arbitrary. The general principle of the Lagrange formulation for free higher-spin field theories in any approach states that the Lagrangian equations of motion must reproduce, as their direct consequences, the conditions defining the irreducible representations of the Poinc\'{a}re group. In the next subsection we will prove that the equation of motion (\ref{eqQ}) is completely consistent with this general principle.

Equation (\ref{eqQ}) possesses a large gauge freedom. Due to nilpotency of the BRST charge in subspace of vectors \p{Phi-s}, the vector $|\Phi_s\rangle$ in equation (\ref{eqQ}) is defined up to gauge transformation
\begin{eqnarray}
|\Phi'_s\rangle \ = \ |\Phi_s\rangle \ + \ Q\,|\Lambda_s\rangle \,,
\label{gauge transf}
\end{eqnarray}
where the gauge parameter $|\Lambda_s\rangle$ with ghost number $-1$
is the extended Fock space vector of the form
\begin{eqnarray}\label{gtQ}
|\Lambda_s\rangle \ = \ \mathcal{P}^+|\hat\lambda_{s-1}\rangle\,.
\label{gauge parameter}
\end{eqnarray}
Let us emphasize that the vectors $|\phi\rangle$, $|\phi_1\rangle$, $|\phi_2\rangle$ in relation (\ref{Phi-s}) are given by the expansions \p{GFState-g}, \p{GFState-g1}, \p{GFState-g-1}, \p{GFState-g-2}. The Fock vector parameter $|\hat\lambda\rangle$ is defined by a similar expansion
\be \label{gauge-par-ex}
|\hat\lambda_{s-1}\rangle=\sum_{k=0}^{s-1}\frac{m^k}{\sqrt{k!}}\,(b^+)^k
|\lambda_{s-1-k}\rangle\,,\qquad
|\lambda_{\sigma}\rangle=\frac{1}{\sigma!}\,\lambda^{\dot{\beta}(\sigma)}_{\alpha(\sigma)}(x)\,c^{\alpha(\sigma)}\,\bar{c}_{\dot{\beta}(\sigma)}|0\rangle
\ee

The equation of motion \p{eqQ} can be identically rewritten as a system of equations for a triplet of vectors
$|\phi_s\rangle,\, |\phi_{1(s-1)}\rangle,\, |\phi_{2(s-2)}\rangle$:
\begin{eqnarray}
&&
\ell_0|\phi_s\rangle-\hat\ell^+|\phi_{1(s-1)}\rangle
=0\,,
\label{236}
\\[5pt]
&&{}
\hat\ell\,|\phi_s\rangle-\hat\ell^+|\phi_{2(s-2)}\rangle+K|\phi_{1(s-1)}\rangle
=0\,,
\label{237}
\\ [5pt]
&&{}
\ell_0|\phi_{2(s-2)}\rangle-\hat\ell\,|\phi_{1(s-1)}\rangle
=0.
\label{238}
\end{eqnarray}
Analogously, the gauge transformation \p{gauge transf} can be identically rewritten as a system of gauge transformations for the above triplet
\be
\delta|\phi_s\rangle \ = \ \hat\ell^+\,|\hat\lambda_{s-1}\rangle\,,
\qquad
\delta|\phi_{1(s-1)}\rangle \ = \ \ell_0\,|\hat\lambda_{s-1}\rangle\,,
\qquad
\delta|\phi_{2(s-2)}\rangle \ = \ \hat\ell\,|\hat\lambda_{s-1}\rangle.
\label{GTAdS0}
\ee
Equations (\ref{236}-\ref{238}) and gauge transformations (\ref{GTAdS0}) are another equivalent form of the equations of motion and gauge transformations for free bosonic massive higher-spin field theory where we have moved from the description in the extended Fock space to the description in the conventional Fock space without ghost operators.

Now it is easy to see that the Lagrangian corresponding to the equation of motion \p{eqQ} is written as follows:
\be
{\cal L}_s
\ = \
\int d\eta_0\; \langle\bar\Phi_s|\,Q\,|\Phi_s\rangle
\,.
\label{actionQ}
\ee
It is evident that this Lagrangian is invariant under the gauge transformation (\ref{gauge transf}).

In the Lagrangian (\ref{actionQ}) one can calculate the scalar product over ghost variables. As a result, we arrive at the equivalent master Lagrangian in terms of
a triplet of vectors $|\phi_s\rangle$, $|\phi_{1(s-1)}\rangle$, $|\phi_{2(s-2)}\rangle$
\begin{eqnarray}
\mathcal{L}_s
&=&
\langle\bar\phi_s|\ell_0|\phi_s\rangle
-\langle\bar\phi_{1(s-1)}|K|\phi_{1(s-1)}\rangle
- \langle\bar\phi_{2(s-2)}|\ell_0|\phi_{2(s-2)}\rangle
\label{Lagr-vector}
\\ [5pt]
&&{}
-\langle\bar\phi_{1(s-1)}|\,\hat\ell\,|\phi_s\rangle
- \langle\bar\phi_s|\,\hat\ell^+|\phi_{1(s-1)}\rangle
+\langle\bar\phi_{2(s-2)}|\,\hat\ell\,|\phi_{1(s-1)}\rangle
+ \langle\bar\phi_{1(s-1)}|\,\hat\ell^+|\phi_{2(s-2)}\rangle\,.
\nonumber
\end{eqnarray}
The triplet of vectors $|\phi_s\rangle$, $|\phi_{1(s-1)}\rangle$, $|\phi_{2(s-2)}\rangle$  in this Lagrangian describes the traceless spin-tensor fields corresponding to massive spins from $0$ to $s$, $s-1$, $s-2$, respectively. 
As will be shown in the next section, after eliminating auxiliary and gauge fields, the physical fields describe single massive spin $s$.
In the limit $m \rightarrow 0$ the dependence on all the operators $b$ and $b^+$
vanishes, and we arrive at the Lagrangian of the massless higher-spin theory.

\subsection{Completion of the Lagrangian formulation}
We proceed to prove that relations (\ref{THEequations}), which define the irreducible representation of the Poinc\'are group with a given mass and spin, are an identical consequence of the equation of motion \p{eqQ} or an identical consequence of the equivalent equations of motion (\ref{236}), (\ref{237}), (\ref{238}).

First of all we note that the vector $|\phi_{1(s-1)}\rangle$ is purely auxiliary and can be eliminated algebraically with the help of the equation of motion \p{237}:
\be
|\phi_{1(s-1)}\rangle={}-K^{-1}\left(\hat\ell\,|\phi_s\rangle-\hat\ell^+|\phi_{2(s-2)}\rangle\right).
\label{1-solut}
\ee
Here we take into account that the operator $K$ (\ref{KNbN}) is non-singular and the inverse operator $K^{-1}$ is well defined (at least in terms of power series).
Substituting the result (\ref{1-solut}) into Lagrangian (\ref{Lagr-vector}), one gets the Lagrangian in terms of vectors $|\phi_s\rangle$ and $|\phi_{2(s-2)}\rangle$ in the form
\begin{eqnarray}
\mathcal{L}_s
&=&
\langle\bar\phi_s|\left(\ell_0 + \hat\ell^+ K^{-1} \hat\ell\right)|\phi_s\rangle
- \langle\bar\phi_{2(s-2)}|\left(\ell_0 - \hat\ell K^{-1} \hat\ell^+\right)|\phi_{2(s-2)}\rangle
\label{Lagr-vector-1}
\\ [5pt]
&&{}
-\langle\bar\phi_{2(s-2)}|\,\hat\ell K^{-1} \hat\ell\,|\phi_s\rangle
- \langle\bar\phi_s|\,\hat\ell^+ K^{-1} \hat\ell^+\,|\phi_{2(s-2)}\rangle \,,
\nonumber
\end{eqnarray}
The corresponding equations of motion are written as follows:
\begin{eqnarray}
&&
\left(\ell_0 + \hat\ell^+ K^{-1} \hat\ell\right)|\phi_s\rangle \ - \ \hat\ell^+ K^{-1} \hat\ell^+\,|\phi_{2(s-2)}\rangle \
= \ 0 \,,
\label{eqs-s}
\\ [5pt]
&&{}
\left(\ell_0 - \hat\ell K^{-1} \hat\ell^+\right)|\phi_{2(s-2)}\rangle \ + \
\hat\ell K^{-1} \hat\ell\,|\phi_{s}\rangle
\ =\ 0\,.
\label{eqs-2s}
\end{eqnarray}
Solution (\ref{1-solut}) for $|\phi_{1(s-1)}\rangle$ preserves gauge invariance; therefore, both Lagrangian (\ref{Lagr-vector-1}) and equations of motion
(\ref{eqs-s}) and (\ref{eqs-2s}) are still gauge invariant under the transformations for the remaining vectors
\be
\delta|\phi_s\rangle \ = \ \hat\ell^+\,|\hat\lambda_{s-1}\rangle\,,
\qquad
\delta|\phi_{2(s-2)}\rangle \ = \ \hat\ell\,|\hat\lambda_{s-1}\rangle.
\label{g-trans-s2s}
\ee

Relations (\ref{g-trans-s2s}) lead to important consequences. Using relations \p{GFState-g} and \p{gauge-par-ex}, where the vectors
$|\phi_s\rangle$ and $|\hat\lambda_{s-1}\rangle$ expressed as polynomials in $b^{+}$, one gets the gauge transformations for the coefficient vectors in these polynomials in the form
\bea
\delta|\varphi_{s-k}\rangle & = & \sqrt{k(2s-k+1)}\,|\lambda_{s-k}\rangle + l^+\,|\lambda_{s-1-k}\rangle\,,
\qquad k=0\div (s-1) \,,
\label{g-trans-s-com}
\\ [7pt]
\delta|\varphi_{0}\rangle & = & \sqrt{s(s+1)}\,|\lambda_{0}\rangle\,.
\label{g-trans-s-com1}
\eea
It is evident that the first term in the right-hand side of (\ref{g-trans-s-com}) vanishes at $k=0$.
Relation (\ref{g-trans-s-com1}) shows that the vector $|\varphi_{0}\rangle$ can be gauged away and the parameter $|\lambda_{0}\rangle$ will be fixed.
Putting $k=s-1$ in (\ref{g-trans-s-com}), one gets
\be
\delta |\varphi_{1}\rangle \ = \ \sqrt{(s-1)(s+2)}\,|\lambda_{1}\rangle + l^{+}\,|\lambda_{0}\rangle\,\
\ee
where the parameter $|\lambda_{0}\rangle$ is already fixed. Using the freedom in choosing the parameter $|\lambda_{1}\rangle$, we can fix this parameter and gauge away the vector $|\varphi_{1}\rangle$. Continuing such a procedure step by step, one gets
\be
\delta |\varphi_{s-1}\rangle \ = \sqrt{2s}\,|\lambda_{s-1}\rangle + l^{+}\,|\lambda_{s-2}\rangle\, .
\ee
The parameter $|\lambda_{s-2}\rangle$ is fixed at the previous step. Hence fixing the parameter $|\lambda_{s-1}\rangle$, we can gauge away the vector $|\varphi_{s-1}\rangle$. At the last step one puts $k=0$. This leads to
\be
\delta |\varphi_{s}\rangle \ = \ l^{+}\,|\lambda_{s-1}\rangle\,.
\label{last step}
\ee
The parameter $|\lambda_{s-1}\rangle$ is already fixed at the previous step. Therefore, relation (\ref{last step}) does not allow one to gauge away the vector $|\varphi_{s}\rangle$ and it remains unchanged. As a result, relations (\ref{g-trans-s-com}) and (\ref{g-trans-s-com1}) provide the possibility of the gauge
\be
|\varphi_{s-k}\rangle \ = \ 0\,,
\qquad k=1\div s \,.
\label{g-fix-s-com}
\ee
In this gauge, the vector $|\phi_s\rangle$ does not depend on $b^+$ and is reduced only to a single vector
\be
|\phi_s\rangle \ = \ |\varphi_{s}\rangle  \,.
\label{phi-varphi}
\ee
The gauge freedom in the theory under consideration is completely fixed and we arrive at non-gauge theory. Nevertheless, we still have a large number of auxiliary fields. Now we discuss how all these auxiliary fields are eliminated and how the fundamental conditions (\ref{THEequations}) are derived.

After imposing the gauge (\ref{g-fix-s-com}), the equations of motion \p{eqs-s}, \p{eqs-2s} take the form
\begin{eqnarray}
&&|E_s\rangle \ := \
\left(\ell_0 + \hat\ell^+ K^{-1} l\right)|\varphi_s\rangle \ - \ \hat\ell^+ K^{-1} \hat\ell^+\,|\phi_{2(s-2)}\rangle \
= \ 0 \,,
\label{eqs-s1g}
\\ [5pt]
&&|E_{2(s-2)}\rangle \ := \ {}
\left(\ell_0 - \hat\ell K^{-1} \hat\ell^+\right)|\phi_{2(s-2)}\rangle \ + \
l K^{-1} l\,|\varphi_{s}\rangle
\ =\ 0\,.
\label{eqs-2s1g}
\end{eqnarray}
Here it has already been used that $b\,|\varphi_s\rangle=0$ and
$\hat\ell\,|\varphi_s\rangle=l\,|\varphi_s\rangle$. Taking into account the form of constraint $\hat\ell^{+}= {l}^{+}+
m b^{+} \big(2s- \mathrm{n}\big)^{1/2}$ in equation (\ref{eqs-s1g}), we see that the first term in this equation contains term independent of $b^{+}$ and terms linear in $b^{+}$. As to the second term in this equation, it contains the terms with $(b^+)^k$, $k=2\div s$. This is easily seen if one takes into account the expansion \p{GFState-g-2} for the vector $|\phi_{2(s-2)}\rangle$.
Let us act on the equation $|E_s\rangle =0$ (\ref{eqs-s1g}) by the operator $(b)^s$ and substitute $|\phi_{2(s-2)}\rangle$ in the form (\ref{GFState-g-2}). Taking into account that acting on the first term in $|E_s\rangle$ leads to identities $b^{s}|\varphi_s\rangle = 0$ or $b^{s-1}|\varphi_s\rangle = 0$.\footnote{Since $|\varphi_s\rangle$ does not contain the operators $b^{+}$, we can either immediately transfer the operator $b^{s}$ to vacuum and get zero or recommute the operator
$b^{s}$ with the operator $b^+$ to obtain $b^{s-1}$ acting on vacuum that gives zero.}  After that, the equation $b^{s}|E_s\rangle =0$ is reduced to
\be
b^{s}\Big( \mathcal{A}+\mathcal{B}\,b^{+}
+ \mathcal{C}(b^{+})^2\Big)\sum_{k=0}^{s-2}\frac{m^k}{\sqrt{k!}}
\, (b^+)^k
|\varphi_{2(s-2-k)}\rangle \ = \ 0\,,
\label{bs}
 \ee
where $\mathcal{A}$, $\mathcal{B}$, $\mathcal{C}$ are some operators. The explicit form for $\mathcal{A}$, $\mathcal{B}$ is unessential for further consideration. It is only important that the operator $\mathcal{C}$ equals
$\mathcal{C}=m^2 K^{-1}\big(2s+1- \mathrm{n}\big)^{1/2}\big(2s+2- \mathrm{n}\big)^{1/2}$ and is nonsingular on all $|\varphi_{2(s-2-k)}\rangle$.
Further, we should recommute the operator $b^{s}$ to $|\varphi_{2(s-2-k)}\rangle$ using \p{b-com} and \p{A-b-com}. The only nonzero result in $b^{s}|E_s\rangle$ arises from the term $b^{s}(b^{+})^s|\varphi_{2(0)}\rangle$, which must vanish due to $b^{s}|E_s\rangle =0$. This means that $|\varphi_{2(0)}\rangle=0$. Putting this relation into
$|\phi_{2(s-2)}\rangle$  in (\ref{eqs-s1g}), acting on this equation by the operator $b^{s-1}$ and repeating the same consideration as before, one gets $|\varphi_{2(1)}\rangle=0$. Then acting by the operator $b^{s-2}$ on equation (\ref{eqs-s1g}), one obtains $|\varphi_{2(2)}\rangle=0$ and so all. At the end,
acting by the operator $b^{2}$ on equation (\ref{eqs-s1g}), one obtains $|\varphi_{2(s-2)}\rangle=0$. As a result, we get  $|\varphi_{2(s-2-k)}\rangle=0$, $k=0\div (s-2)$ in expression \p{GFState-g-2} for $|\phi_{2(s-2)}\rangle$.
This is a compact form of the conditions
\be \label{phi-2-comp-0}
\varphi{}_{2,\,}{}^{\dot\beta(s-2-k)}_{\alpha(s-2-k)}=0\,,
\qquad k=0\div (s-2)
\ee
on the Singh-Hagen-like auxiliary component fields (see (\ref{f-2s})).
Hence we have
\be \label{phi-2-0}
|\phi_{2(s-2)}\rangle=0
\ee
in the equation $|E_s\rangle =0$ (\ref{eqs-s1g}). After that, this equation takes the form
\be
\left(l_{0} + \hat\ell^{+}K^{-1}\ell\right)|\varphi_{s}\rangle \ = \ 0\,.
 \label{eqs-s1g-s}
\ee
Acting on this equation by the operator $b$, one gets
\be \label{eqs-s1g-s-1}
l\,|\varphi_s\rangle=0\,.
\ee
Then, the conditions \p{eqs-s1g-s} and \p{eqs-s1g-s-1} lead to
\be \label{eqs-s1g-s-0}
l_0\,|\varphi_s\rangle=0\,.
\ee
Equations (\ref{eqs-s1g-s-1}) and (\ref{eqs-s1g-s-0}) coincide with conditions (\ref{THEequations})
defining the irreducible representation of the
Poinc\'are group with a given mass and spin. We emphasize that these conditions are direct consequences of the equation of motion (\ref{eqQ})
which are coding basic equations \p{condition1} for component fields.
In essence, the latter circumstance serves as a justification of the correctness and consistency of the BRST approach to the Lagrangian formulation of free massive higher-spin field theory.

Note that when equations \p{phi-2-0}, (\ref{eqs-s1g-s-1}) and (\ref{eqs-s1g-s-0}) are satisfied, equation \p{eqs-2s1g} is satisfied automatically.

It is clear that the Lagrangian \p{Lagr-vector-1} can be simplified by imposing the gauge \p{phi-2-0}. After that, the Zinoviev-like gauge auxiliary vector
$|\phi_{1(s-2)}\rangle$ is eliminated, and we arrive with the Lagrangian in terms a doublet of Fock space  vectors $|\varphi_{s}\rangle$ and $|\phi_{2(s)}\rangle$ in the form
\begin{eqnarray}
\mathcal{L}_s
&=&
\langle\bar\varphi_s|\left(\ell_0 + l^+ K^{-1} l\right)|\varphi_s\rangle
- \langle\bar\phi_{2(s-2)}|\left(\ell_0 - \hat\ell K^{-1} \hat\ell^+\right)|\phi_{2(s-2)}\rangle
\label{Lagr-vector-1-s}
\\ [5pt]
&&{}
-\langle\bar\phi_{2(s-2)}|\,l K^{-1} l\,|\varphi_s\rangle
- \langle\bar\varphi_s|\,l^+ K^{-1} l^+\,|\phi_{2(s-2)}\rangle \,.
\nonumber
\end{eqnarray}
The Lagrangian (\ref{Lagr-vector-1-s}) now contains only basic higher-spin field and Singh-Hagen-like auxiliary fields.

We emphasize that there is no need to specially prove the equivalence of all such different Lagrangian formulations. All of them are based on the same single criterion of correctness: the equations of motion reproduce the relations that determine the irreducible representations of the Poinc\'are group with a given mass and spin (see, e.g., \cite{Buch3,Buch4}).

\subsection{Field contents}

We have shown that the BRST Lagrangian description of massive arbitrary spin fields is formulated in terms of triplet vectors
$|\phi_{s}\rangle$, $|\phi_{1(s-1)}\rangle$ and $|\phi_{2(s-2)}\rangle$. Each of these vectors includes the standard space-time fields the number of which increases with spin. Let us discuss the structure of these vectors in more detail. The expansions \p{GFState-g}, \p{GFState-g1}, (\ref{GFState-g-1}), (\ref{GFState-g-2})
allow one to describe completely the structure of component fields in the form
\begin{eqnarray}
&& |\phi_{s}\rangle\ : \qquad\ \ \Big\{ |\varphi_{s}\rangle \,,
\underbrace{|\varphi_{s-1}\rangle \,,|\varphi_{s-2}\rangle \,,\
\ldots\,, \ |\varphi_{1}\rangle \,, |\varphi_{0}\rangle }_{\mbox{Zinoviev-like gauge states}} \Big\}\,, \lb{0s}\\
&& |\phi_{1(s-1)}\rangle \ : \quad \Big\{
\underbrace{|\varphi_{1(s-1)}\rangle \,, |\varphi_{1(s-2)}\rangle \,,\
\ldots\,, \
|\varphi_{1(1)}\rangle \,, |\varphi_{1(0)}\rangle  }_{\mbox{BRST auxiliary states}} \Big\}\,, \lb{1s} \\
&& |\phi_{2(s-2)}\rangle \ : \quad \Big\{
\underbrace{|\varphi_{2(s-2)}\rangle \,, |\varphi_{2(s-3)}\rangle  \,,\
\ldots\,, \ |\varphi_{2(1)}\rangle \,, |\varphi_{2(0)}\rangle
 }_{\mbox{Singh-Hagen-like auxiliary states}}   \Big\} \lb{2s}
\end{eqnarray}
with spins $0\div s$, $0\div (s-1)$, and $0\div (s-2)$, respectively.
These sets of vectors are born from the vacuum by acting on the powers of the additional creation operator $b^+$
inherent only in the massive case.
The vectors \p{0s}, \p{1s}, \p{2s} are described by the fields
\begin{eqnarray}
&& \varphi{}^{\dot\beta(s-k)}_{\alpha(s-k)}\,,\ k=0\div s \ : \qquad\qquad\ \ \  \Big\{ \varphi^{\dot\beta(s)}_{\alpha(s)} \,,
\underbrace{\varphi{}^{\dot\beta(s-1)}_{\alpha(s-1)} \,,\varphi{}^{\dot\beta(s-2)}_{\alpha(s-2)} \,,\
\ldots\,, \ \varphi{}^{\dot\beta}_{\alpha} \,, \varphi_{0} }_{\mbox{Zinoviev-like gauge fields}} \Big\}\,, \lb{f-0s} \\
&& \varphi{}_{1,\,}{}^{\dot\beta(s-1-k)}_{\alpha(s-1-k)}\,,\ k=0\div (s-1) \ : \quad \Big\{
\underbrace{\varphi{}_{1,\,}{}^{\dot\beta(s-1)}_{\alpha(s-1)} \,, \varphi{}_{1,\,}{}^{\dot\beta(s-2)}_{\alpha(s-2)} \,,\
\ldots\,, \ \varphi{}_{1,\,}{}^{\dot\beta}_{\alpha} \,, \varphi{}_{1,\,0}}_{\mbox{BRST auxiliary fields}} \Big\}\,, \lb{f-1s} \\
&& \varphi{}_{2,\,}{}^{\dot\beta(s-2-k)}_{\alpha(s-2-k)}\,,\ k=0\div (s-2)\ : \quad \Big\{
\underbrace{\varphi{}_{2,\,}{}^{\dot\beta(s-2)}_{\alpha(s-2)} \,, \varphi{}_{2,\,}{}^{\dot\beta(s-3)}_{\alpha(s-3)} \,,\
\ldots\,, \ \varphi{}_{2,\,}{}^{\dot\beta}_{\alpha} \,, \varphi{}_{2,\,0} }_{\mbox{Singh-Hagen-like auxiliary fields}}   \Big\}\,. \lb{f-2s}
\end{eqnarray}
In the BRST approach, the vectors  \p{0s}, \p{1s}, \p{2s} and the corresponding fields \p{f-0s}, \p{f-1s}, \p{f-2s}
play a very clear and understandable role.

The vectors $|\phi_{1(s-1-k)}\rangle$ and the corresponding fields
$\varphi{}_{1,\,}{}^{\dot\beta(s-1-k)}_{\alpha(s-1-k)}(x)$, $k=0\div (s-1)$ are auxiliary
and are eliminated algebraically by using their equations of motion \p{1-solut}.
They present in the initial action as a result of using the BRST triplet formulation of the theory.

The Lagrangian \p{Lagr-vector-1} obtained after eliminating $|\phi_{1(s-1-k)}\rangle$
possesses  gauge symmetry and contains gauge St\"{u}ckelberg degrees of freedom for the massive higher-spin theory.
This Lagrangian corresponds to the massive spin theory proposed by Zinoviev in \cite{Zin1,Zin2,Zin3}.
If we take into account the  gauge transformations \p{g-trans-s-com}, \p{g-trans-s-com1},
then it is natural to consider the vectors $|\varphi_{s-k}\rangle$ and the corresponding fields
$\varphi{}^{\dot\beta(s-k)}_{\alpha(s-k)}(x)$, $k=0\div (s-1)$ as St\"{u}ckelberg vectors/fields.
After gauge fixing \p{g-fix-s-com} that excludes these vectors/fields,
we obtain the Lagrangian in the Singh-Hagen formulation \cite{Sing1} in terms of the vectors
$|\varphi_{s}\rangle$, $|\varphi_{2(s-2-k)}\rangle$ and the corresponding fields
$\varphi^{\dot\beta(s)}_{\alpha(s)}(x)$, $\varphi{}_{2,\,}{}^{\dot\beta(s-2-k)}_{\alpha(s-2-k)}(x)$, $k=0\div (s-2)$.
In this formulation the fields $\varphi{}_{2,\,}{}^{\dot\beta(s-2-k)}_{\alpha(s-2-k)}(x)$, $k=0\div (s-2)$
are auxiliary whereas the field $\varphi^{\dot\beta(s)}_{\alpha(s)}(x)$ is physical one.

Note that the scheme considered here is a generalization of the massless case (see for example \cite{BFIK-24}).
Using the mass parameter in the expansions \p{GFState-g}, \p{GFState-g-1}, \p{GFState-g-2}, we have only vectors
$|\phi_{s}\rangle=|\varphi_{s}\rangle$, $|\phi_{1(s-1)}\rangle=|\varphi_{1(s-1)}\rangle$,
$|\phi_{2(s-2)}\rangle=|\varphi_{2(s-2)}\rangle$
remaining at $m{=}\,0$ and
we completely reproduce the triplet formulation of a  massless particle of an arbitrary spin (helicity).

\setcounter{equation}0
\section{Summary}

In this paper we have significantly modified and simplified the previously developed BRST construction to derive the Lagrangian formulation for the free massive integer higher-spin field in four-dimensional flat space-time. We emphasize the novelty of the approach under consideration.

Unlike the previous approaches \cite{Buch1, Buch2}, where the Lagrangian BRST construction was developed for massive higher-spin fields,  we use the formulation in terms in the Fock space vectors formed by annihilation and creation operators with two-component undotted and dotted spinor indices.\footnote{Such a Fock space was introduced earlier to describe the supersymmetric massless higher-spin fields in \cite{BKout}.} Therefore, the corresponding massive higher-spin fields are irreducible spin-tensors with respect to the Lorentz group. This leads to a significant simplification of the whole BRST construction since now there is no need to introduce into the BRST charge the constraint associated with tracelessness. The matter is that such spin-tensor fields will automatically be traceless after restoration of vector indices (see, e.g., \p{twocomp}).\footnote{The Lagrangian formulation of 4d massive higher spin fields in terms of two-component spinors was considered in \cite{DELP3} without using the BRST construction.}

As well as in the previous approaches, the conditions defining the irreducible representation of the Poinc\'are group with a given mass and spin are formulated in the  Fock space under consideration with the help of the constraints $l_0$ and $l$ (\ref{l_0}) according to conditions (\ref{THEequations}). The constraint $l_0$ is the Klein-Gordon operator and provides the mass irreducibility while the constraint $l$ is the transversality operator and provides the spin irreducibility.
It was shown that the constraint $l$ and conjugate constraint $l^+$ are the second class ones,
which leads to a problem of constructing the Hermitian BRST charge. The transformation of the second class constraints into the first class ones is based on the conversion procedure (see, e.g., \cite{FSh, BF}). In our case, the conversion was realized by introducing the new bosonic annihilation and creation operators $b, \, b^+$. This makes it possible to construct an equivalent theory which is characterized by modified constraints \p{op-dyn} that form the first class algebra \p{THEequations2b} in the Fock space. Moreover, the action of the operators $b$ and $b^+$ on the vacuum, unlike the massless case, generates additional fields (see, e.g., \p{GFState-g}) which can be associated with auxiliary fields in the massive higher-spin theory.
The BRST charge \p{Q(r)} constructed on the basis of modified first class constraints defines the gauge invariant Lagrangian and generates a triplet field theory. The formal construction turns out to be marvelously similar to the massless higher-spin field theory that is also formulated in terms of the fields triplet. However, since in the massive case we have additional annihilation and creation operators $b, \, b^+$  in the Fock space, we get additional auxiliary fields in comparison with the massless case.

The main issue of any approach to the Lagrangian formulation for free higher-spin fields is the proof that the Lagrange equations of motion reproduce as their direct consequences the relations that determine the irreducible higher-spin representations of the Poinc\'{a}re group. Such a proof in the original paper \cite{Sing1} (see also \cite{KUZ1}) looked like extremely complicated for an arbitrary spin $s$. In the previous papers on the BRST approach to massive higher-spins
\cite{Buch1, Buch2, Buch3} the study of this issue involved the consideration of a large number of specific details and also was not very simple. In the present paper, we have developed an elegant and extremely simple derivation of the above relations from the Lagrange equations of motion. The procedure uses only very well-known properties  $[b,\,b^{+}]=1$ and $b|0\rangle = 0$ of the bosonic annihilation and creation operators.

The Lagrangian was derived in terms of expectation values over the Fock space vectors, in fact in terms of vacuum expectation values from the products of some numbers of annihilation and creation operators. After calculating the above expectation values, one gets the Lagrangian in terms of spin-tensor space-time fields.
These fields are naturally divided into basic higher-spin field, Zinoviev-like auxiliary fields, Singh-Hagen-like auxiliary fields and specific BRST-like auxiliary fields. Eliminating parts of the auxiliary fields with the help of gauge fixing or equations of motion, we have the formally different but equivalent Lagrangian formulations. Their equivalence does not require a special test; it is evident because all of them are based on the same relations determining the irreducible representation of the Poinc\'are group with a given mass and spin.

It is interesting to note once more that formally the same triplet Lagrangian structure takes place in both massless and in massive cases. The only difference is the number of auxiliary fields and the form of constraints with their same algebra on the Fock space under consideration. Therefore, one can hope to derive interaction vertices for the massive higher-spin field theory using the vertices for the massless one.

Note that infinite and finite higher-spin (helicity) field theories in the framework of the BRST formulation on the basis of oscillators with two-component spinor indices and the corresponding two-component spin-tensor fields were earlier considered in \cite{Buch5} and \cite{BFIK-24}. In this paper, we have studied the last earlier unconsidered case of the BRST Lagrangian formulation in terms of two-component spin-tensors,
viz. the case of the BRST formulation for massive higher-spin fields in four dimensions.

\renewcommand\theequation{A.\arabic{equation}} \setcounter{equation}0
\section*{Appendix \ \ Component field Lagrangian}

After using the relations
\be\label{norm-vect}
\langle0|\,a_{\alpha_1}\ldots a_{\alpha_s} c^{\beta_1} \ldots c^{\beta_s} |0\rangle=
s!\,\delta^{\beta_1}_{(\alpha_1}\ldots \delta^{\beta_s}_{\alpha_s)}\,,\qquad
\langle0|\,a_{\alpha(s_1)} c^{\beta(s_2)} |0\rangle=0\ \  \mbox{при}  \ \  s_1 \neq s_2 \,,
\ee
\be\label{norm-vect-bar}
\langle0|\,\bar a^{\dot\alpha_1}\ldots \bar a^{\dot\alpha_s} \bar c_{\dot\beta_1} \ldots \bar c_{\dot\beta_s} |0\rangle=
s!\,\delta_{\dot\beta_1}^{(\dot\alpha_1}\ldots \delta_{\dot\beta_s}^{\dot\alpha_s)} \,,\qquad
\langle0|\,\bar a^{\dot\alpha(s_1)} \bar c_{\dot\beta(s_2)} |0\rangle=0 \ \  \mbox{при}  \ \  s_1 \neq s_2 \,,
\ee
\be\label{norm-vect-b}
\langle0|\,(b)^{k} (b^+)^{k} |0\rangle= k!\,,\qquad
\langle0|\,(b)^{k_1} (b^+)^{k_2} |0\rangle=0
\ \  \mbox{при}  \ \  k_1 \neq k_2
\ee
and expansions
\p{GFState-g}, \p{GFState-g1}, \p{GFState-g-a}, \p{GFState-g1-a}, the master BRST Lagrangian \p{Lagr-vector}
yields the following component Lagrangian:
\begin{eqnarray}
\mathcal{L}_s
&=&
\sum\limits_{k=0}^{s}m^{2k} \bar\varphi_{\dot\beta(s-k)}^{\alpha(s-k)}\left(\Box-m^2\right)\varphi^{\dot\beta(s-k)}_{\alpha(s-k)}
\ - \ 2\sum\limits_{k=0}^{s-1}m^{2k} (s-k) \bar\varphi{}_{1,\,}{}_{\dot\beta(s-1-k)}^{\alpha(s-1-k)} \, \varphi{}_{1,\,}{}^{\dot\beta(s-1-k)}_{\alpha(s-1-k)}
\label{Lagr-comp}
\\ [5pt]
&&{}
\nonumber
-\sum\limits_{k=0}^{s-2}m^{2k} \bar\varphi{}_{2,\,}{}_{\dot\beta(s-2-k)}^{\alpha(s-2-k)}
\left(\Box-m^2\right)\varphi{}_{2,\,}{}^{\dot\beta(s-2-k)}_{\alpha(s-2-k)}
\\ [5pt]
&&{}
+ \sum\limits_{k=0}^{s-1}m^{2k} (s-k) \Big( \bar\varphi{}_{1,\,}{}_{\dot\beta(s-1-k)}^{\alpha(s-1-k)} \,\partial_{\dot\beta}^{\alpha} \varphi{}^{\dot\beta(s-k)}_{\alpha(s-k)}  \ +  \
\partial^{\dot\beta}_{\alpha} \bar\varphi{}_{\dot\beta(s-k)}^{\alpha(s-k)}\,\varphi{}_{1,\,}{}^{\dot\beta(s-1-k)}_{\alpha(s-1-k)} \Big)
\nonumber
\\ [5pt]
&&{}
- \sum\limits_{k=0}^{s-1}m^{2k+2} \sqrt{(2s-k)(k+1)}
\Big(\bar\varphi{}_{1,\,}{}_{\dot\beta(s-1-k)}^{\alpha(s-1-k)} \, \varphi{}^{\dot\beta(s-1-k)}_{\alpha(s-1-k)} \ + \
\bar\varphi{}_{\dot\beta(s-1-k)}^{\alpha(s-1-k)} \, \varphi{}_{1,\,}{}^{\dot\beta(s-1-k)}_{\alpha(s-1-k)}\Big)
\nonumber
\\ [5pt]
&&{}
+ \sum\limits_{k=0}^{s-1}m^{2k} (s-1-k) \Big(
\bar\varphi{}_{1,\,}{}_{\dot\beta(s-1-k)}^{\alpha(s-1-k)} \,\partial^{\dot\beta}_{\alpha} \varphi{}_{2,\,}{}^{\dot\beta(s-2-k)}_{\alpha(s-2-k)}
\ +  \ \partial_{\dot\beta}^{\alpha} \bar\varphi{}_{2,\,}{}_{\dot\beta(s-2-k)}^{\alpha(s-2-k)} \,\varphi{}_{1,\,}{}^{\dot\beta(s-1-k)}_{\alpha(s-1-k)} \Big)
\nonumber
\\ [5pt]
&&{}
+ \sum\limits_{k=0}^{s-1}m^{2k} \sqrt{k(2s+1-k)} \Big(
\bar\varphi{}_{1,\,}{}_{\dot\beta(s-1-k)}^{\alpha(s-1-k)} \, \varphi{}_{2,\,}{}^{\dot\beta(s-1-k)}_{\alpha(s-1-k)}
 \  +  \
\bar\varphi{}_{2,\,}{}_{\dot\beta(s-1-k)}^{\alpha(s-1-k)} \, \varphi{}_{1,\,}{}^{\dot\beta(s-1-k)}_{\alpha(s-1-k)}
\Big)
\,.
\nonumber
\end{eqnarray}
The field model with the Lagrangian \p{Lagr-comp} is described by the triplet of fields:
$\varphi^{\dot\beta(s-k)}_{\alpha(s-k)}$, $k=0\div s$,
$\varphi{}_{1,\,}{}^{\dot\beta(s-1-k_1)}_{\alpha(s-1-k_1)}$, $k_1=0\div (s-1)$ and
$\varphi{}_{2,\,}{}^{\dot\beta(s-2-k_2)}_{\alpha(s-2-k_2)}$, $k_2=0\div (s-2)$
having the mass dimensions
$$
[\varphi^{\dot\beta(s-k)}_{\alpha(s-k)}]=[\varphi{}_{2,\,}{}^{\dot\beta(s-2-k)}_{\alpha(s-2-k)}]=[m]^{1-k}\,,\qquad
[\varphi{}_{1,\,}{}^{\dot\beta(s-1-k)}_{\alpha(s-1-k)}]=[m]^{2-k}\,.
$$
Note that the penultimate term in  \p{Lagr-comp} does not contain a term with $k=s-1$, since for this value of $k$ there is no field $\varphi{}_{2,\,}{}^{\dot\beta(s-2-k)}_{\alpha(s-2-k)}$. This is taken into account automatically, since the coefficient in this term for $k=s-1$ is identically equal to zero. Similarly, the last term does not contain a term with $k=0$, since for this value of $k$ there is no field $\varphi{}_{2,\,}{}^{\dot\beta(s-1-k)}_{\alpha(s-1-k)}$. This is also taken into account automatically, since the coefficient in this term for $k=0$ is identically equal to zero. It is instructive to compare a simple and compact Lagrangian (\ref{Lagr-vector}) written in terms of the Fock space vectors with an equivalent cumbersome Lagrangian (\ref{Lagr-comp}) written in terms of component fields.

For $m{=}0$, only terms with three fields $\varphi^{\dot\beta(s)}_{\alpha(s)}$, $\varphi{}_{1,\,}{}^{\dot\beta(s-1)}_{\alpha(s-1)}$, $\varphi{}_{2,\,}{}^{\dot\beta(s-2)}_{\alpha(s-2)}$ survive in the Lagrangian \p{Lagr-comp}, and the Lagrangian \p{Lagr-comp} goes over to the Fronsdal Lagrangian for the massless spin (helicity) field $s$ in the triplet BRST formulation presented in \cite{BFIK-24}.


\begin{thebibliography}{96}

\bibitem{revVas}
M.A.\,Vasiliev, \textit{Higher spin gauge theories in any
dimension}, Comptes Rendus Physique, 5 (2004) 1101,
{\tt arXiv:hep-th/0409260}.

\bibitem{revBCIV}
X.\,Bekaert, S.\,Cnockaert, C.\,Iazeolla, M.A.\,Vasiliev, \textit{Nonlinear
higher spin theories in various dimensions}, in Higher spin gauge
theories: Proceedings, 1st Solvay Workshop : Brussels, Belgium,
12-14 May, 2004, 132-197, {\tt arXiv:hep-th/0503128}.

\bibitem{revFT}
A.\,Fotopoulos, M.\,Tsulaia, \textit{Gauge invariant Lagrangians for free and interacting higher spin fields, A review of the BRST formulation,}
Int.\,J.\,Mod.\,Phys. A \textbf{24} (2008) 1, {\tt arXiv:0805.1346\,[hep-th]}.

\bibitem{revBek}
X.\,Bekaert, N.\,Boulanger, P.\,Sundell, \textit{How higher spin gravity surpasses
the spin two barrier: no-go theorems versus yes-go examples}, Rev.\,Mod.\,Phys.
\textbf{84} (2012) 987, {\tt arXiv:1007.0435\,[hep-th]}.

\bibitem{DidSk}
V.E.\,Didenko, E.D.\,Skvortsov, \textit{Elements of Vasiliev theory,}
{\tt arXiv:1401.2975 [hep-th]}.

\bibitem{reviewsV}
M.A.\,Vasiliev, \textit{Higher-spin theory and space-time metamorphoses,}
Lect. Notes Phys. 892 (2015) 227, {\tt arXiv:1404.1948 [hep-th]}.

\bibitem{revBekSk}
X.\,Bekaert, E.D.\,Skvortsov, \textit{Elementary particles with continuous spin}, Int.\,J.\,Mod.\,Phys. A \textbf{32} (2017) 1730019, {\tt arXiv:1708.0103 [hep-th]}.

\bibitem{Snowmass}
X.\,Bekaert, N.\,Boulanger, A.\,Campaneoli, M.\,Chodaroli, D.\,Francia,
M.\,Grigoriev, E.\,Sezgin, E.\,Skvortsov, \textit{Snowmass White Paper:
Higher spin gravity and higher spin symmetry}, {\tt
arXiv:2205.01567\,[hep-th]}.

\bibitem{Ponomarev}
D.\,Ponomarev, \textit{Basic introduction to higher-spin theories,}
{\tt arXiv:2206.15385\,[hep-th]}.


\bibitem{OCHI}
A.\,Ochirov, E.\,Skvortsov, \textit{Chiral approach to massive higher spins}, Phys.~Rev.~Lett. {\bf 129} (2022) 125027,
{\tt arXiv:2207.14597\,[hep-th]}.

\bibitem{Lin}
L.W.\,Lindwasser, \textit{Consistent actions for massive particles interacting with electromagnetism and gravity,} JHEP \textbf{08} (2024) 081, {\tt arXiv:2309.03901\,[hep-th]}.

\bibitem{KUZ1}
J.H.\,Fegebank, S.M.\,Kuzenko, \textit{Quantisation of the gauge-invariant models for massive higher-spin bosonic fields}, {\tt arXiv:2310.00951\,[hep-th]}.

\bibitem{SkTs}
E.\,Skvortsov, M.\,Tsulaia,
\textit{Cubic action for spinning black holes from massive higher-spin gauge symmetry},
JHEP \textbf{02} (2024) 202, {\tt arXiv:2312.08184\,[hep-th]}.

\bibitem{KUZ2}
A.J.\,Fegebank, S.M.\,Kuzenko, \textit{Equivalence of gauge-invariant models for massive integer-spin fields}, Phys.\,Rev. D {\bf 110} (2024) 10, 105014, {\tt arXiv:2406.02573\,[hep-th]}.

\bibitem{DELP1}
W.\,Delplanque, E.\,Skvortsov, \textit{Symmetric vs. chiral approaches to massive fields with spin}, Class.\,Quant.\,Grav. {\bf 41} (2024) 24, 245018,
{\tt arXiv:2405.13706\,[hep-th]}.

\bibitem{DELP2}
W.\,Delplanque, E.\,Skvortsov, \textit{Massive spin three-half field in a constant electromagnetic background}, JHEP {\bf 08} (2024) 173,
{\tt arXiv:2406.14148\,[hep-th]}.

\bibitem{DELP3}
W.\,Delplanque,
{\it All actions for free massive higher-spin fields},
Phys.\,Rev.  {\bf D111} (2025) 4, {\tt arXiv:2411.03463\,[hep-th]}.

\bibitem{Sing1}
L.F.S.\,Singh, C.R.\,Hagen, \textit{Lagrangian formulation for arbitrary spin 1. The bosonic case}, Phys.\,Rev. D {\bf 9} (1974) 898.

\bibitem{Sing2}
L.F.S.\,Singh, C.R.\,Hagen, \textit{Lagrangian formulation for arbitrary spin 2. The fermionic case}, Phys.\,Rev. D {\bf 9} (1974) 910.

\bibitem{Schwinger}
J.\,Schwinger, \textit{Particles, Sources and Fields}, Volume\,1, 426\,pp., Addison-Wesley Publishing Company, 1970.

\bibitem{FP}
M.\,Fierz and W.\,Pauli, \textit{On relativistic wave equations for particles of arbitrary spin in an electromagnetic field}, Proc.\,Roy.\,Soc.\,Lond.\, A \textbf{173} (1939) 211.

\bibitem{Zin1}
Yu.M.\,Zinoviev,
\textit{Gauge invariant description of massive high spin particle},
Serpukhov,Institute of High Energy Physics, report number: IFVE-83-91,1983.

\bibitem{Zin2}
S.M.\,Klishevich, Yu.M.\,Zinovev,
\textit{On electromagnetic interaction of massive spin-2 particle},
Phys. Atom. Nucl. {\bf 61} (1998) 1527, {\tt arXiv:hep-th/9708150}.

\bibitem{Zin3}
Yu.M.\,Zinoviev,
\textit{On massive higher-spin particles in AdS}, {\tt arXiv:hep-th/0108192}.

\bibitem{Metsaev1}
R.R.\,Metsaev, \textit{Massive totally symmetric fields in AdS(d)}, Phys.\,Lett.\, B \textbf{590} (2004) 95, {\tt arXiv:hep-th/0312297\,[hep-th]}.

\bibitem{Metsaev2}
R.R.\,Metsaev, \textit{Gauge invariant formulation of massive totally symmetric fermionic fields in (A)dS space,} Phys.\,Lett.\, B \textbf{643} (2006) 205,  {\tt arXiv:hep-th/0609029\,[hep-th]}.


\bibitem{Buch1}
I.L.\,Buchbinder, V.A.\,Krykhtin,
\textit{Gauge invariant Lagrangian construction for massive bosonic higher spin
fields in D dimensions},
Nucl.\,Phys.\,  B \textbf{727} (2005) 537, {\tt arXiv:hep-th/0505092}.

\bibitem{Buch2}
I.L.\,Buchbinder, V.A.\,Krykhtin, P.M.\,Lavrov,
\textit{Gauge invariant Lagrangian formulation of higher spin massive bosonic field theory in AdS space},
Nucl.\,Phys.\, B \textbf{762} (2006) 386, {\tt arXiv:hep-th/0608005}.

\bibitem{Buch3} 
I.L.\,Buchbinder, V.A.\,Krykhtin, H.\,Takata,
\textit{Gauge invariant Lagrangian construction for massive bosonic
mixed symmetry higher spin fields},
Phys. Lett.  B {\bf656} (2007) 253, {\tt arXiv:0707.2181\,[hep-th]}.

\bibitem{Buch4}
I.L.\,Buchbinder, A.V.\,Galajinsky, \textit{Quartet unconstrained formulation for massive higher spin fields,}\, JHEP \textbf{11} (2008) 081,
{\tt arXiv:0810.2852\,[hep-th]}.

\bibitem{Pash}
A.I.\,Pashnev,
\textit{Composite systems and field theory for a free Regge trajectory},
Theor. Math. Phys. \textbf{78} (1989) 272.

\bibitem{T1}
A.\,Pashnev, M.\,Tsulaia, \textit{Dimensional reduction and the BRST approach to the description of a Regge trajectory,} Mod.\,Phys.\,Lett.\, A \textbf{12} (1997) 861, {\tt arXiv:hep-th/9703010}.

\bibitem{T2}
X.\,Bekaert, I.L.\,Buchbinder, A.\,Pashnev, M.\,Tsulai, \textit{On Higher Spin Theory: Strings, BRST, Dimensional Reduction,}, Class.\,Quant.\,Grav.\, \textbf{21} (2004) S1457, {\tt arXiv:hep-th/03122252}.

\bibitem{Vas25}
M.A.\,Vasiliev, \textit{Supersymmetric higher-spin gauge theories in any $d$ and their coupling constants within BRST formalism}, {\tt arXiv:2503.10967\,[hep-th]}.

\bibitem{Zin4}
M.V.\,Khabarov, Yu.M.\,Zinoviev, \textit{Massive higher spin fields in the frame-like multispinor formalism,} Nucl.\,Phys.\, B \textbf{948} (2019) 11477, {\tt arXiv:1906.03438\,[hep-th]}.

\bibitem{BKout}
I.L.\,Buchbinder, K.\,Koutrolikos, \textit{BRST analysis of the supersymmetric higher spin field models}, JHEP \textbf{12} (2015) 106, {\tt arXiv:1510.06569\,[hep-th]}.

\bibitem{BFIK-24}
I.L.\,Buchbinder, S.A.\,Fedoruk, A.P.\,Isaev, V.A.\,Krykhtin,
\textit{On BRST Lagrangian formulation of massless higher spin fields},
Russ.\,Phys.\,J.\, \textbf{67} (2024) 1806, {\tt arXiv:2412.08298\,[hep-th]}.

\bibitem{BK}
I.L.\,Buchbinder, S.V.\,Kuzenko, \textit{Ideas and methods of supersymmetry
and supergravity, or a walk through superspace}, IOP Publishing, Bristol and Philadelphia, 1995;
Revised Edition, 1998.


\bibitem{FSh}
L.D.\,Faddeev, S.L.\,Shatashvili, \textit{Realization of the Schwinger term in the Gauss law and the possibility of correct quantization of a theory with anomalies}, Phys.\,Lett. B \textbf{167} (1986) 225.

\bibitem{BF}
I.A.\,Batalin, E.S.\,Fradkin, \textit{Operatorial quantization of dynamical systems subject to second class constraints}, Nucl.\,Phys. B \textbf{279} (1987) 514.

\bibitem{Buch5}
I.L.\,Buchbinder, V.A.\,Krykhtin, H.\,Takata, \textit{BRST approach to Lagrangian construction for bosonic continuous spin field},
Phys.\,Lett. B \textbf{785} (2018) 315, {\tt arXiv:1806.01640\,[hep-th]}.


\end{thebibliography}
\end{document}